# Regulatory Considerations for Using Artificial Intelligence Models to Reduce Sample Sizes in Registrational Studies

Aaron M. Smith[1], Tala Fakhouri[2], Run Zhuang[1], Jonathan R. Walsh[1]
[1]Unlearn.AI, [2]Parexel International

## Abstract

Applications of artificial intelligence (AI) in drug development continue to increase at a rapid pace. Regulatory authorities have provided increasingly clear perspectives on the use of AI in regulated applications, including recent draft guidance from FDA that provides a 7-step risk-based framework to assess AI model credibility for these cases. We present an application of AI models to prospectively reduce the planned sample size in a randomized controlled trial, using model-derived prognostic covariates. This can shorten trial timelines, enable faster decision making, and lower costs. When treatments are effective and tolerable they can be accessible to patients sooner, which is a compelling use case for the FDA guidance. We walk through each of the steps in the guidance, providing general recommendations for model development, evaluation, and approaches for sample size determination, with the intent of providing a clear set of guidelines on how to engage with the FDA guidance and advance responsible use of AI in drug development. We demonstrate the application with an example in Alzheimer’s Disease.

## Introduction

Enrollment burden is a persistent challenge for many clinical trials, with the inability to meet enrollment targets, one of the most cited causes of clinical trial failures [1]. Various trial elements can help reduce the sample size of clinical trials, including innovative trial designs, more sensitive endpoints, and broader patient populations. Each of these examples faces headwinds and may not be practical for a large fraction of registrational trials. However, the goal of reducing sample size in trials, while maintaining adequate power to demonstrate substantial evidence of efficacy and safety in the intended population remains important, as it can lead to faster clinical trials that bring effective treatments to patients earlier as well as reduce the patient burden for trials in ineffective treatments.

Recently, an approach called prognostic covariate adjustment (PROCOVA) was introduced that can reduce the required sample size for a randomized controlled trial (RCT) by utilizing the framework of covariate adjustment [2]. Most statistical analyses of RCTs include adjustments for baseline covariates, which are variables measured pre-randomization that characterize participants. Covariate adjustment is generally well understood and well discussed in regulatory guidance, with FDA recommending that sponsors adjust for variables that are expected to be prognostic or otherwise associated with outcomes of interest in analyses of trial endpoints [3]. Typically, covariates are incorporated from stratification or because they are believed to be prognostic, meaning they predict the outcome regardless of treatment (e.g., disease severity, comorbidities, etc.) [4, 5].

As shown in Figure 1, PROCOVA uses a model, such as an AI model, to create a prognostic covariate for one or more endpoints in a clinical trial. For each participant, the prognostic covariate is generated from baseline data and is the predicted value for the endpoint being analyzed, assuming that participant was randomized to the control arm. The model is trained, tuned, and tested on historical data from similar populations to predict one or more endpoints that will be measured in the trial.

**1. Model Development**

training

training / tuning data

model

**2. Model Testing**

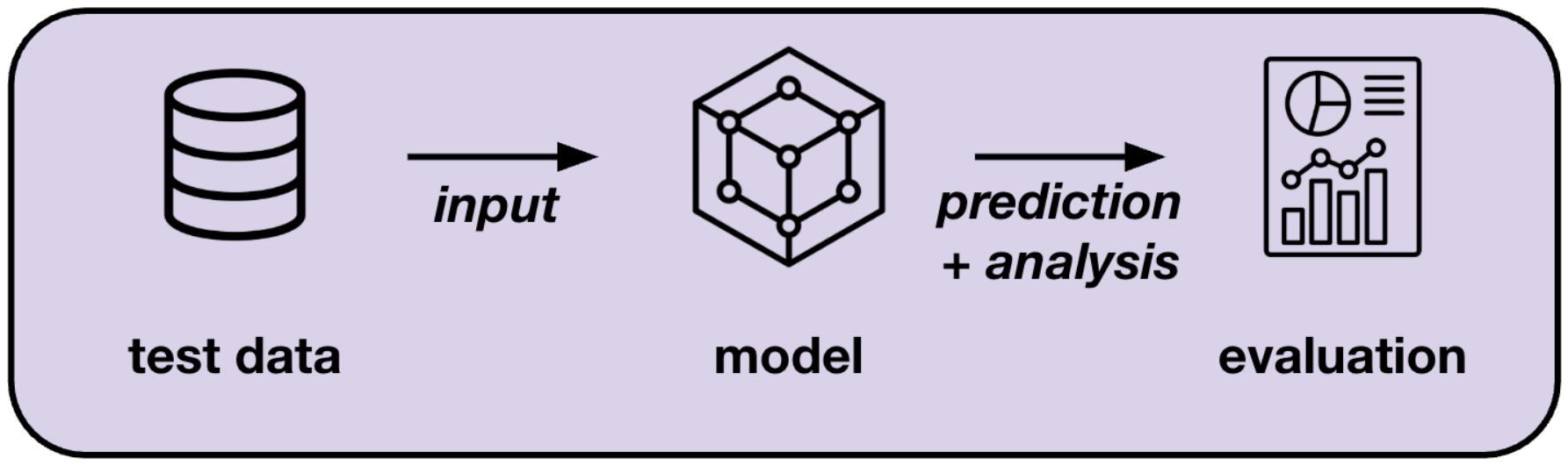

**3. Prognostic Variable Creation**

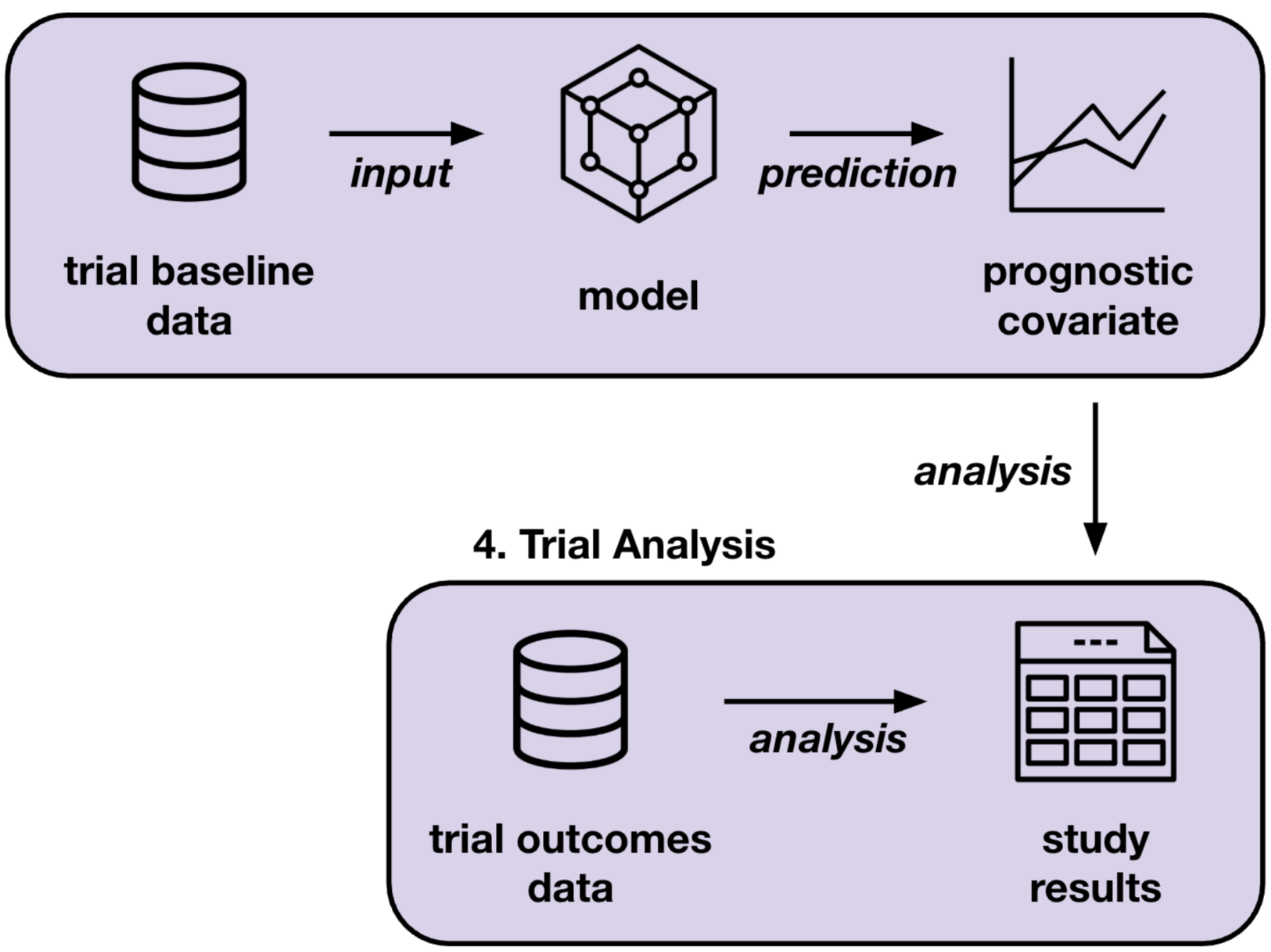

**Figure 1:** Key steps in the use of PROCOVA. Models are trained by training and tuning them on historical data from similar populations (Step 1). Models are evaluated on test data to determine performance, which is used to evaluate the adequacy for the context of use (Step 2). In a clinical trial, prognostic variables are created from the model by inputting baseline data from each trial participant into the model and obtaining predictions (Step 3). At the end of the study, those predictions plus the trial outcomes are used to analyze efficacy outcomes (Step 4). Note that steps 3 and 4 are performed to prospective test model performance, for example to evaluate the performance in a prospective phase 2 trial ahead of planned use to reduce sample size in a phase 3 trial.

PROCOVA can be used to reduce the sample size in a clinical trial by prospectively incorporating the expected prognostic benefit from the model-derived covariate for relevant endpoints in the sample size calculation [6]. This requires a clear expectation for the amount of prognostic benefit that the model-derived covariate will provide, as overestimation of benefit can lead to reductions in power. This implies careful considerations around the development and validation of the prognostic model.

In early 2025, the U.S. Food and Drug Administration (FDA) released draft guidance outlining a 7-step framework for evaluating the credibility of artificial intelligence (AI) in clinical trials [7]. The guidance takes a risk-based approach, emphasizing the importance of context of use (COU)—how and where AI is applied—and encouraging appropriate safeguards based on that context. This guidance is a crucial and large step forward in articulating a framework for the development and use of AI in regulatory decision making, enabling actionable implementation and the development of AI applications in clinical trials. While the position of regulators continues to evolve [8], carefully reviewing and incorporating the principles outlined in this guidance can provide important lessons to developers of relevant AI applications. As companies invest large capital, time, and resources into adopting AI models into their global drug development programs, joint principles between agencies become increasingly important for providing common interpretation of how regulators intent to apply these risk-based frameworks [9].

In this paper, we apply the 7-step risk-based framework to assess model credibility for the general case of sample size reduction using model-derived covariates from PROCOVA. We discuss key factors that influence model risks, including strategies that can mitigate risks. We provide recommendations for model development, evaluation, and processes for selecting the amount of sample size reduction. The application of a prognostic model in Alzheimer's Disease is used as practical example. Presented as a step-by-step walkthrough of the 7-step risk-based framework, this paper is intended as a guide for model developers, regulatory experts, and clinical development teams to realize this valuable application of AI in clinical trials. Supplementary materials provide additional guides and discussion.

## Reducing Sample Sizes in Clinical Trials with PROCOVA

Covariate adjustment in RCTs is an ideal framework to apply AI models in clinical research. The benefits of randomization ensure several important properties for covariate adjusted analyses in standard use cases, such as unbiased treatment effects estimation, type I error rate control, and additional gains of statistical power [3]. The prognostic benefit from covariates, whether model-derived or not, is realized by yielding treatment effect estimates with better precision (i.e., smaller variance) and therefore improving statistical power. The fact that standard applications of covariate adjustment preserve the type I error rate is a key factor that regulators consider when evaluating a new drug application. Practitioners must still exercise caution when dealing with complex analysis models and scenarios, such as very small studies, the case of predictive covariates (i.e., those that modify treatment effect) in mixed models with repeated measures (MMRM) analyses, or the use of non-collapsible estimands in non-continuous endpoints. These scenarios may require additional statistical expertise and testing because the standard assumptions or asymptotic properties in more common use cases may not hold, requiring greater care to avoid biased results or incorrect conclusions about treatment efficacy.

PROCOVA is designed to maximize the benefits of covariate adjustment by providing a highly prognostic single covariate, with the AI model learning how to fully use baseline data to robustly predict the expected control outcome. Under a set of mild assumptions [2], the model-derived covariate can be shown to be the most powerful single covariate that can be determined, and it consolidates prognostic information across baseline data. This makes it a persuasive approach to realize sample size reduction in clinical trials.

In this work, we focus on the case of continuous endpoints, as these are discussed most thoroughly in the FDA guidance and have the most straightforward statistical properties [3]. For other analysis models or endpoint types that allow covariate adjustment, a version incorporating a model-derived covariate can also be defined [10, 11, 12, 13, 14, 15]. Special considerations may arise with other types of endpoints or more complex statistical analysis models, particularly for non-collapsible estimands or when covariates are predictive. In addition, issues of collinearity require careful attention, especially for AI-derived prognostic scores, which often share information with baseline variables. While detailed strategies for managing these challenges are beyond the scope of this paper, the commentary in this paper can be considered as an interpretation of the FDA guidance, illustrated with examples specific to continuous outcomes.

The FDA guidance recommends that sponsors define the context of use, estimate AI model risk, and describe the model development plan which includes descriptions of training, tunning, and testing data, model description and evaluation, and emphasizing the importance of early engagement [7]. We will focus on the model development and credibility assessment steps rather than the operational work of incorporating the model-derived covariate into a study; while important, that is beyond the scope of the discussion here. We note that during trial conduct, the model-derived covariate is incorporated into the analysis datasets prior to database lock and into statistical analyses like any other covariate, leaving the study conduct unaffected other than recognizing a smaller sample size.

## Overview of the FDA's Seven-Step Risk-Based Framework for AI Model Credibility

The FDA's credibility framework provides a structured, risk-based approach to evaluating AI applications in clinical trials. The first four steps place the AI application into context, evaluate risks, and define a plan to evaluate the credibility of the model, as shown in Table 1.

**Table 1:** The steps in FDA's 7-step risk-based framework, derived from the recent draft guidance [7].

| Step | Purpose |
|---|---|
| 1 | Define the Question of Interest: Clearly articulate the specific decision or regulatory question the AI model aims to address. |
| 2 | Define the Context of Use (COU): Outline the role and scope of the AI model within drug development, including how its outputs will influence study design or decision-making. |
| 3 | Assess AI Model Risk: Evaluate the potential impact of model error, considering both its influence on decisions and the consequences of those decisions. |
| 4 | Develop a Plan to Establish AI Model Credibility Within the COU: Describe the model development and evaluation process in relation to the intended application. |
| 5 | Execute the Plan |
| 6 | Document the Results and Discuss Any Deviations |

| 7 | Determine the Adequacy of the AI Model for the COU |
|---|---|

These steps are carried out before the use of a model in the intended application as part of model development and evaluation, although there may be actions that take place during the study to validate the proper use of the model.

Importantly, the FDA guidance suggests early engagement with regulators around the model credibility assessment to align on activities in the assessment, identify potential challenges, and how those challenges may be addressed; these engagements should be used to provide the information in the credibility assessment to elicit input. Early engagement may be valuable – even when model development is ongoing – since the question of interest, context of use, and model risk assessment can often be clarified and aligned on with the FDA without a developed model.

## Applying the Seven-Step Framework to Sample Size Reduction via PROCOVA

### Step 1: Define the Question of Interest

As stated in the guidance, the question of interest should "describe the specific question, decision, or concern being addressed by the AI model". In this case, the AI model is being used with PROCOVA to improve statistical efficiency, with the goal of prospectively applying that efficiency to reduce the sample size while maintaining the prespecified power for the study. The central question, then, concerns power for the study: "Does the proposed trial design, which incorporates PROCOVA-based covariate adjustment, maintain adequate power at a reduced sample size compared with an unadjusted design?".

This is a narrow but important question for efficacy analyses, especially for sponsors, and the key endpoints include the primary endpoint and any secondary endpoints considered when powering the study. An incorrect decision on the question of interest can impact the ability to demonstrate a statistically significant benefit and gain approval for an effective therapy.

Note that the question of interest is not confined to the AI model but instead applies to the overall problem addressed by it. In this case, the question of adequate power also concerns the estimate of endpoint variance, dropout rate, and any other factors in the design that influence statistical power. The narrowness of the question of interest is due to the fact that PROCOVA fits within a standard analysis framework of covariate adjustment; this would not be the case with other trial design approaches such as historical borrowing.

### Step 2: Define the Context of Use for the AI Model

The COU is vital as it defines the role of the AI model in answering the question of interest. It determines the parameters and scope for evaluating model risk, delineating how the model and other non-modeling information will be used in answering the question of interest.

The context of use is the following. The AI model will be used in the trial design to create prognostic covariates for a set of endpoints in the study. These covariates will be used in efficacy analyses to improve statistical efficiency by increasing the precision of treatment effect estimates. The sample size calculation for the study will account for the expected increase in precision from the covariate to lower the required sample size and will be informed by the prognostic value of the model-derived covariate on data from previous clinical trials, observational studies or registries, or real-world data. The target patient population, endpoints, and other features of the planned trial design should be provided along with the context of use.

Robust estimation of other factors (principally the endpoint variance, dropout rate, assumed treatment effect size, and other elements of the statistical analysis model such as the effect of imputation on endpoint variance or the covariance structure in repeated measures) influencing the sample size calculation will be used to ensure adequate powering of the design, independent of the prognostic covariates, in answering the question of interest.

## Step 3: Assess the AI Model Risk

The FDA 7-step risk-based guidance categorizes risk along two key dimensions: decision consequence and model influence. Decision consequence reflects the potential adverse impact of an incorrect decision on the question of interest, evaluated broadly and not specific to the AI model itself. A high decision consequence is one that has a direct and significant impact on patient safety, product quality, or the reliability of study results. Model influence is the role of the model relative to other factors in answering the question of interest, and is evaluated with respect to patient safety, product quality, and reliability of study results. A highly influential model is one that plays a proportionally large role in answering the question of interest.

These dimensions guide the development of a tailored credibility assessment plan (Step 4). FDA engagement at these stages is iterative; the agency may suggest refinements to the model credibility plan based on the model's risk profile, context of use, and the question of interest. Aligning early on this plan – through actions such as integration into statistical analysis plans, protocols, in questions to FDA, during meetings – lays the groundwork for a more productive regulatory review process.

The decision consequence for the question of interest is low. Incorrect decisions on whether the study is adequately powered can lead to a study that has a lower probability of demonstrating a statistically significant outcome but does not jeopardize the study's validity (i.e. whether the study design, conduct, and analysis are scientifically sound and unbiased) and safety protections remain intact. Running a study with smaller power is a risk borne by the sponsor, and sponsors must decide how strongly to power studies to ensure that effective treatments are shown to be statistically significant. We note that in practice, the potential reductions in power from contemplated sample size decreases are at the scale where the ethical obligations to run a well-powered study are still met.

The influence of the model on the question of interest can range from low to high depending on the amount of sample size reduction enabled by PROCOVA and the associated risk for reducing power. If the reduction is small, then its contribution to the power reduction risk is low and the model influence is low. If other factors that influence study power are well controlled and the amount of sample size reduction is large, then the model influence is high.

The question of risk then becomes whether the effective sample size is sufficiently below the required sample size such that the negative impact on power is meaningful rather than minimal. Different approaches can be used to mitigate this risk as described in the next section.

### *Risk mitigation strategies*

The principal and necessary method to control risk is to validate model performance. Without evaluation, there is no quantitative data on which to support a sample size reduction determination. Data from relevant populations should be used wherever possible to provide additional evidence for model performance, and may be a pre-requisite to use depending on the level of model influence. Model testing should play a prominent role in the model credibility plan (Step 4; see also the Supplementary Information) to ensure robust evaluation.

Two other risk mitigation strategies focus on the context in which the model is applied rather than the development of the model itself: sample size re-estimation as a blinded interim analysis and considerations around how the target power is set in relation to standard covariates.

When an interim analysis will be conducted, blinded sample size re-estimation can be used to significantly mitigate the risk of reducing power [16]. The prognostic performance of the model-derived covariate, as well as standard covariates and the value of other factors influencing the power such as the endpoint variance and dropout rate, can be assessed while maintaining blinding and the effective sample size computed. If the effective sample size falls short of the planned sample size, enrollment can be increased to compensate. In practice, because not all data is available at the interim analysis, the risk cannot be eliminated but the degree and likelihood of power reduction can both be reduced, potentially dramatically. Simulations can be performed to estimate these likelihoods quantitatively.

Additionally, an important consideration is the prognostic value in baseline covariates and stratification factors. If they are expected to be materially prognostic, then this prognostic value may provide protection against significant reductions in power. Standard trial design defines the sample size in terms of a target power for an unadjusted analysis. This means that in practice, there are two power gains to consider: the gain from the standard covariates or stratification factors and the gain from the model-derived covariate. If we believe the power gain from the standard covariates are robust, then that provides an expected minimum power gain over the target power, reducing the risk of losing power when sample size is reduced.

## Steps 4, 5, and 6: Model Credibility Plan, Execution, and Documentation

The model credibility plan, or credibility assessment plan, is the specific document that communicates the question of interest, context of use, model risk and detailed information about the development and evaluation of the model used to evaluate the adequacy of the model for the context of use. This document should be created in line with GxP requirements and is provided to regulators, such as a part of a formal meeting package or through another engagement channel and provides a way to receive interactive feedback on the model risk and credibility assessment. Regulatory engagement is often useful after the plan has been developed in Step 4, but before it is executed in Step 5. Once work moves to execution of the plan there may be practical challenges if feedback motivates changes to the development plan. This should be balanced against the ability to get timely feedback.

The draft guidance contains several recommendations for specific topics to be presented in the credibility assessment plan. These cover the description of the model, including inputs and outputs, architecture, and feature selection process; the data used to train and evaluate the model, including how data was collected, transformed, managed, and its relevance for training and evaluation; the model training process, including the software, learning algorithms, optimization methods, calibration, and specialized techniques such as pre-training or ensembling; the model evaluation process, including the appropriateness and management of the test data, performance metrics, and the rationale for the evaluation methods. Additionally, the credibility assessment plan should describe practices used to ensure proper data management and software practices throughout the development and evaluation lifecycle. The Supplementary Information contains specific recommendations for these steps.

Careful evaluation of the model risk and clear presentation of the credibility plan make it much easier to execute on model development and to evaluate model adequacy for the context of use. Like the credibility plan, we recognize that the development steps and systems may differ between teams and use cases. To evaluate the credibility assessment plan, it is valuable to collect documented evaluation results of the model-derived covariate along with details of the trial design and risk mitigation strategies. Collectively this information helps sponsors determine whether the model is adequate for the context of use, allowing for an easier decision-making process in step 7. We provide detailed recommendations for data, modeling, and evaluation specific to the context of use, as well as execution and documentation of results in the Supplementary Information.

### Step 7: Evaluate Adequacy for the Context of Use

The final step is to weigh the results of the credibility assessment and determine whether sample size will be reduced in the study, and if so by how much. This is not a completely algorithmic process as it requires weighing the information reported in step 6. Across the range of potential sample size reductions, there are two major considerations: the meaningfulness of the reduction and degree of risk for losing power.

For a given magnitude of sample size reduction, two considerations should be weighed. First, the degree of risk for reducing power. This risk can be largely quantified, for example the minimum power expected if the model-derived covariate provides no prognostic value or if it provides a minimum level of prognostic value motivated by evaluation performance and design features of the study. Second, the meaningfulness of the amount of sample size reduction. There may be a minimum for sample size reduction above which is worth the added logistical complexity or cost.

The sample size reduction chosen should balance these considerations. First, it should be determined if meaningful sample size reduction can be supported within tolerable risk levels for reducing power. There may be circumstances where the sample size reductions that can safely be realized are insufficient for the added complexity. Second, the amount of sample size reduction should be chosen, and a written evaluation should be made of the risk of reducing power, both in terms of likelihood and degree. This risk presentation should be expected to be a central point of discussion with regulators.

## An Example in Alzheimer's Disease

In this section we demonstrate the considerations around prospective sample size reduction with PROCOVA, using an example in Alzheimer's Disease (AD). We utilize the results of a previously developed machine learning model of AD progression that was applied to a retrospective analysis of a completed, 24-month phase 2 clinical trial in early AD [17]. We will assume the intended application is prospective sample size reduction for a phase 3 trial, with a focus on evaluating adequacy for the context of use (Step 7 in the risk-based framework).

We consider a hypothetical future phase 3 AD clinical trial, with 1000 participants randomized with 1:1 ratio. The trial is assumed to have co-primary endpoints of the 18-month change score from baseline in Clinical Dementia Rating Sum-of-Boxes (CDR-SB) and the Alzheimer's Disease Assessment Scale Cognitive Subscale 14 (ADAS-Cog 14) powered to 90%. CDR-SB evaluates both cognition and function, whereas ADAS-Cog 14 focuses primarily on cognition. Regulatory guidance requires demonstrating efficacy across both domains [18].

Although this example pre-dates the FDA guidance [7], we enumerate the ideal interaction points with the regulators over the intended use for the phase 3 trial. In this example, the AD model was developed concurrently with the conduct of the trial. The retrospective analysis along with the potential context of use, to apply the AD model to prospective sample size reduction in a future phase 3 trial, were conceived after model development was complete. As discussed above, when possible, it is preferrable to engage regulators when the context of use is well-defined, even if model development is not complete. Once the retrospective analysis on the phase 2 study is finalized, the adequacy for the context of use can be performed. This is another ideal time to engage with regulators, as the credibility assessment is complete and the plan to employ the AD model in the phase 3 application can be defined. Such a meeting can occur at the End of Phase 2 meeting or as a Type C meeting ahead of the phase 3 initiation. Circumstances may change when these interaction points occur, such as when the model is developed and when the credibility assessment is completed. For example, ample evidence may exist ahead of a phase 2 trial that provides sufficient information to determine a planned sample size reduction for the phase 3 trial.

The AD model was evaluated in a retrospective analysis of a completed, 24-month phase 2 trial in early AD [17, 19]. In addition, during development it was tested on the control arms of three AD clinical trials with mild-to-moderate patient population (phase 2 or phase 3 trials, labeled studies A, B, and C and of varying duration between 12 and 24 months) that were held out from training [20, 21, 22]. Variance reduction of the model-derived prognostic covariate for each held out testing dataset is shown in Table 2. In these studies, the baseline covariates provide very little prognostic value, so we ignore them for simplicity. We note that studies A, B, and C use versions of ADAS-Cog with fewer items than the planned study.

**Table 2:** Test data evaluation of the AD model on a recently completed Phase 2 trial in early AD [19], reporting results from [17], as well as three AD clinical trials with mild-to-moderate AD patient population. The variance reduction of the model-derived prognostic covariate for each outcome and timepoint are shown; endpoints are the change from baseline of the outcome at the timepoint in question. Entries without a value are cases where the study duration is shorter than the timepoint.

| Cohort | Outcome | # Participants | Variance Reduction of the Model-Derived Covariate | | |
|---|---|---|---|---|---|
| | | | 12 months | 18 months | 24 months |
| Phase 2 | CDR-SB | 453 | 16.1% | 15.9% | 13.0% |
| Study A control | CDR-SB | 116 | 13.8% | 14.0% | 13.0% |
| Study B control | CDR-SB | 136 | 8.5% | 9.5% | - |
| Study C control | CDR-SB | 46 | 21.1% | - | - |

| Phase 2 | ADAS-Cog 14 | 453 | 8.5% | 4.5% | 9.3% |
|---|---|---|---|---|---|
| Study A control | ADAS-Cog 11 | 93 | 4.3% | 1.8% | 5.2% |
| Study B control | ADAS-Cog 11 | 132 | 10.3% | 12.4% | - |
| Study C control | ADAS-Cog 11 | 46 | 30.9% | - | - |

The phase 2 results are the most relevant for a future phase 3 study in the sample population, with the test studies being more severe in terms of AD progression. Given the strong results on CDR-SB in the phase 2 results, we believe that a sample size reduction equivalent to a 10% variance reduction can be supported. This is equivalent to the reduction of 100 participants from the 1000 participants in a phase 3 study if the overall sample size is reduced and the randomization ratio is maintained. If the sample size is reduced in the control arm starting from a 1:1 randomization, this is equivalent to a 18% control arm reduction. For a case of a planned phase 3 trial with 90% power, even if the model-derived covariate offers *zero* prognostic benefit the power will be above 86% (if the overall sample size is reduced). Therefore, there is a small risk of significant power loss, and with sample size re-estimation performed in the interim analyses the risk of reducing power is minimized.

## Conclusions

Reducing the sample size of clinical trials, especially registrational trials, without sacrificing the quality of evidence is a valuable way to lower costs and shorten timelines in drug development. PROCOVA provides a rigorous and validated methodology to achieve this for randomized trials, and interest in PROCOVA has been continually growing as a result. It offers a unique opportunity that is otherwise difficult to achieve: applying AI models in safe, well understood, and regulatory acceptable ways to achieve enhance statistical power and improve decision making in clinical trials. This particular application focuses on the use of PROCOVA to prospectively reduce sample size while maintaining statistical integrity and compliance with regulatory expectations.

As regulators increasingly articulate their stance on AI in drug development, ensuring alignment with regulatory guidance is important to establish robust processes to AI use in regulatory decision making. This paper provides a detailed discussion and considerations for applying FDA's recent guidance on AI in drug development, specifically focused on the application of reducing sample size through an AI model-derived covariate with PROCOVA.

In the future, we expect that PROCOVA will be increasingly used as part of phase 2 and 3 trials to increase statistical efficiency by adding power and reducing sample size. As studies complete and treatments go through regulatory approval, greater exposition of the steps taken to successfully achieve efficiency gains in clinical trials can aid the field in being able to reproducibly succeed in applying this new technology. Additionally, future directions can include the application of similar methods to more complex designs, such as non-continuous endpoints, Bayesian analyses for RCTs with a prior on the accuracy of the model derived covariate in predicting control outcomes, or open label applications in appropriate settings. The information presented in this paper is intended to be helpful to both sponsors and regulators as PROCOVA and other AI applications are increasingly being used in drug development.

# Supplementary Information

In this section we provide additional discussion on the use of PROCOVA for sample size reduction. This includes a summary of the main body's discussion for each step in FDA's risk-based framework, as well as recommendations for model development to ensure a robust credibility assessment.

## Sample size reduction and power tradeoffs

One important consideration for clinical trial sponsors is how sample size will be reduced, and by how much. Sponsors may elect to be conservative in sample size reduction, capturing gains in statistical power in combination with sample size reductions. How sample size reduction is distributed across arms is also a consideration. For example, all arms can be reduced equally, the control arm size can be reduced, or sample size reallocated to achieve a target randomization ratio (e.g., 2:1). The control arm reduction option preserves sample sizes on treatment to build safety databases and can make the study more attractive to patients as they will be more likely to receive treatment, but in some indications un-equal randomization ratios are disfavored because they can lead to an increase in placebo effects.

As an example, in Figure 2, we show two curves for power versus sample size in a hypothetical study, with and without a prognostic covariate that provides an estimated 15% variance reduction in the treatment effect estimate. In practice, a sponsor may elect to realize a portion of this benefit as sample size reduction and any remaining benefit will lead to a more highly powered study. For example, if the original study design had 1000 participants, factoring the entire expected benefit of the prognostic score into the sample size calculation to maintain the original randomization ratio would yield a study with 850 participants; in practice a sponsor may choose less total sample size, such as 900-950 participants.

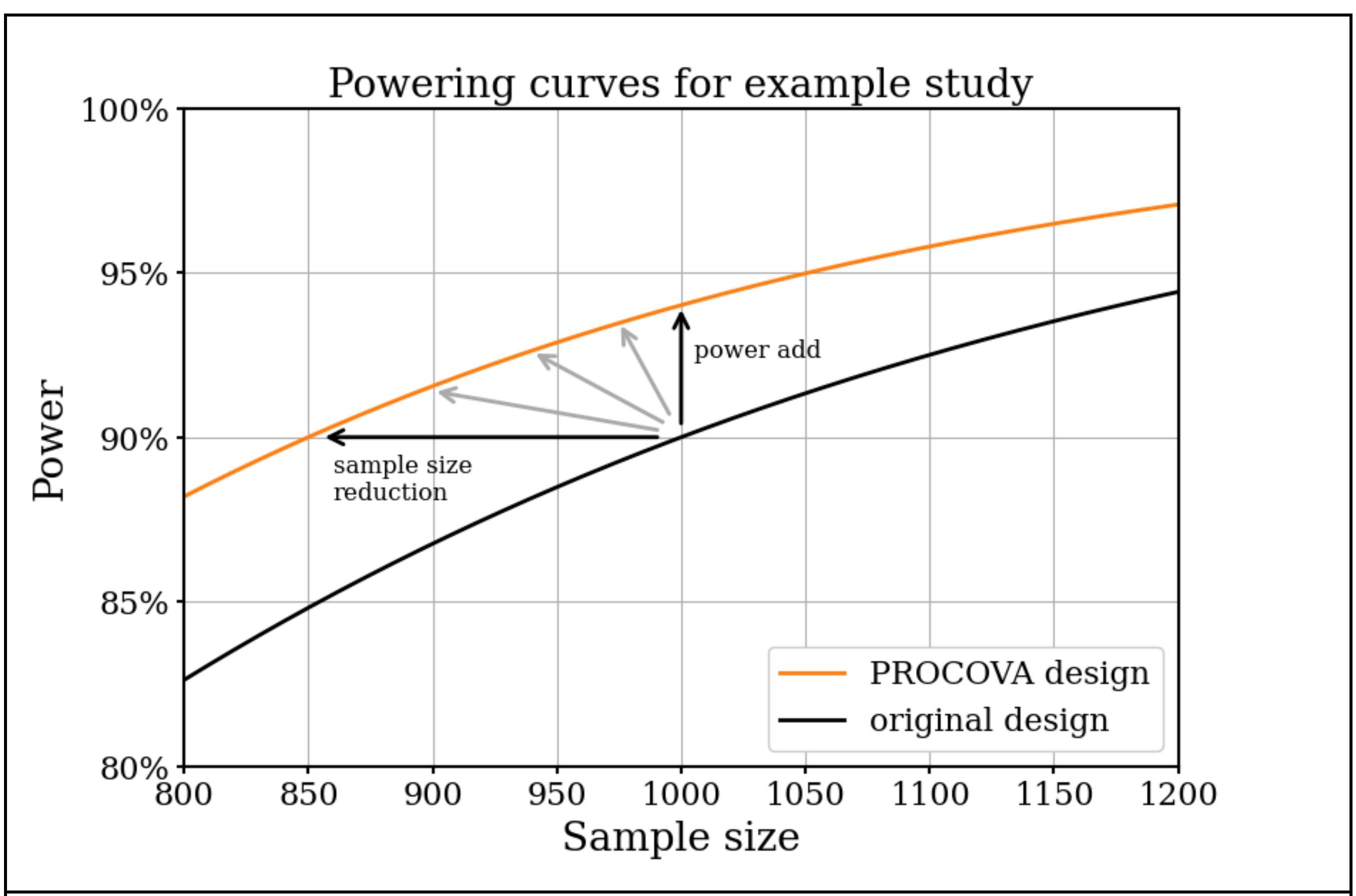

**Figure 2:** Example power curves for a trial where 1000 participants are required for 90% power in a traditional design, and where PROCOVA provides an estimated 15% variance reduction. The benefits can be recognized through entirely sample size reduction, which if the randomization ratio is maintained yields a study with 850 participants and 90% power, or the benefits can be recognized entirely as power gain, which yields a study with 1000 participants and 94% power, or a mixture (represented by the gray arrows). The amount of sample size reduction chosen is informed by a risk assessment for reducing power if the prognostic covariate underperforms and is a goal of the risk assessment.

## Effective sample size to assess risk

For a common trial design with a single continuous primary endpoint using PROCOVA, the power calculation is influenced by several factors. A main driver is the expected treatment effect, which depends on evidence from prior studies on the treatment, domain knowledge, the endpoint, and planned population for the study. The power is defined for a specific treatment effect, so we will regard it as fixed. Others are the expected variance of the endpoint, which depends on the planned population; the expected dropout rate, which depends on the study design and planned population, and study design parameters such as the significance level, which are often fixed. With PROCOVA, the variance of reduction enabled by the model-derived covariate is also a factor, which depends on the expected performance for the planned population. Finally, the sample size in each arm is a key factor, one that is computed based on the other factors to achieve a target power. Unlike the other factors, it is not a feature of the study population but is chosen by the sponsor. We note that standard covariates typically do not factor into the power calculation for a study and are used at analysis time to correct for randomization imbalance and add power.

Power loss is ultimately the result of a sample size that is too small, but it can come from an underestimation of the endpoint variance, dropout rate, or model-derived covariate performance. There is inherently uncertainty in each of these quantities, and typically results from previous, similarly designed trials are used to estimate the endpoint variance and dropout rate. When such trials do not exist, observational or real-world data may be used. In general, there is often non-trivial uncertainty on these quantities.

We can define the *effective sample size* as the sample size we would have computed had we known ahead of the trial the standard deviation, performance of the model-derived covariate, or other study design parameters (e.g., dropout rate) that influence the sample size calculation. Note that these are all observed in the study, and so after a trial is completed such a calculation can be made and compared to the actual sample size used in the trial.

Figure 3 shows the impacts to power as a function of the effective sample size compared to the actual sample size for the study. For example, if the effective sample size is 10% lower than the actual sample size, the power is reduced to 75.7% (80% design power) and 86.8% (90% design power).

We can use this framework to assess the magnitude of impact on power. For example, it is a reasonable question to ask, "if my study is powered to 90%, how problematic is a 3.2% power decrease?". The answer depends on factors like the overall confidence in study design and the initial power, as well as the qualitative risk that such an outcome arises. A decrease of a few percent power from 90% still yields a well-powered study; for a design power of 80% we may feel quite differently about power loss.

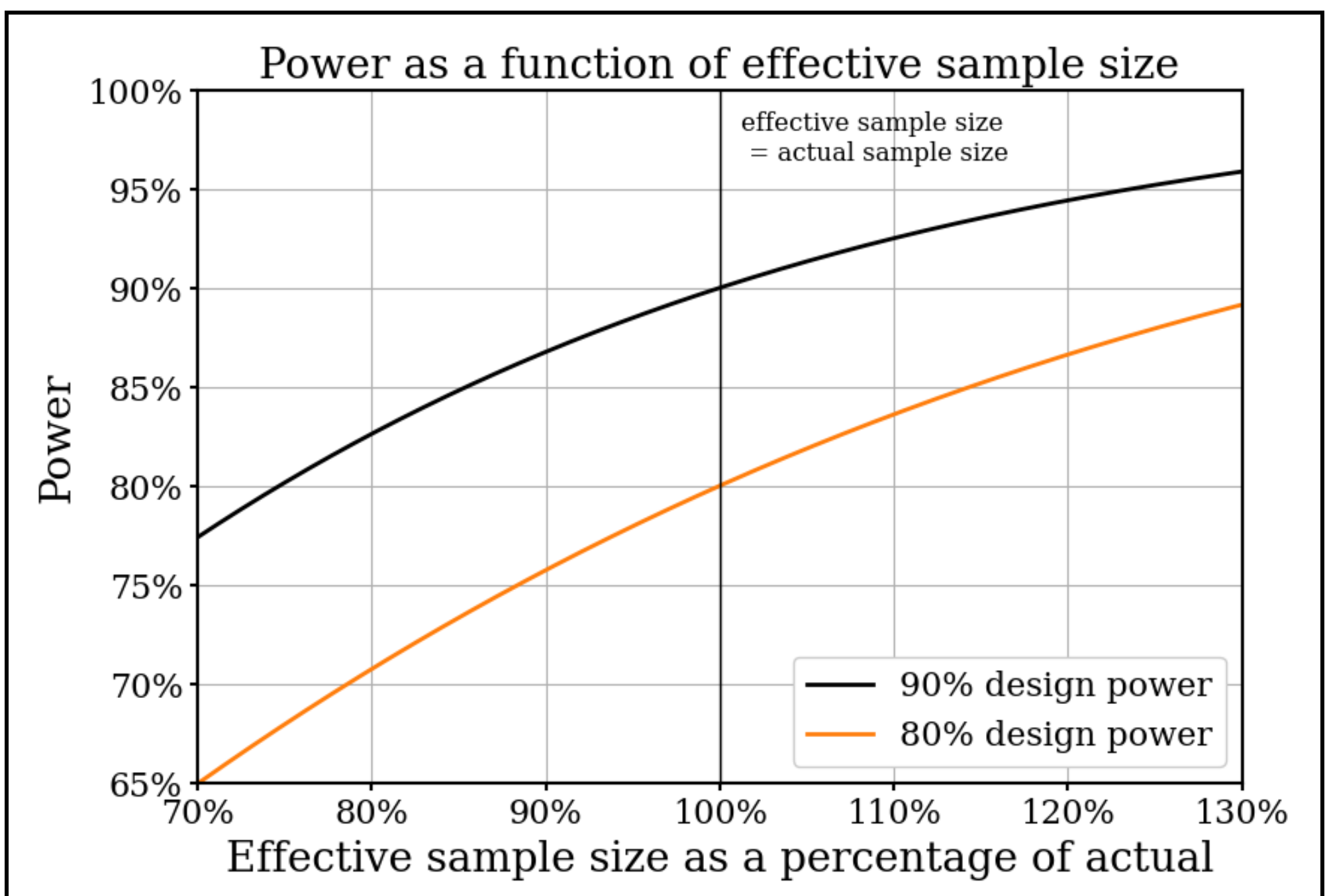

**Figure 3:** The power as a function of the effective sample size of the study, expressed as the difference to the actual sample size. For example, a value of 90% on the x-axis means the effective sample size is 10% lower than the actual sample size. The two curves show the power for either an 80% or 90% design.

We note that there are different ways to realize a reduced sample size for a study. Two typical approaches considered are reducing the size of all arms while maintaining a target randomization or reducing the control arm alone. Reducing the control arm can make enrollment more appealing for prospective participants, since they have a greater chance of receiving the new treatment. In some indications where the enrollment ratio is known or suspected to influence placebo effects, reducing the placebo arm alone may be inadvisable. The size of safety datasets is also a consideration, as reducing the size of the treated population will limit the size of the safety dataset; however, the effects are fairly small until sample size reductions become significant (e.g., greater than 25% of treated participants reduced).

## Summary and Recommendations for Risk-Based Framework Steps

| **Step 1: The Question of Interest** |
| --- |
| Does the proposed trial design, which incorporates PROCOVA-based covariate adjustment, ensure adequate power while minimizing sample size? |

| **Step 2: The Context of Use of the AI Model** |
| --- |
| The AI model will be used in the trial design to create prognostic covariates for a set of endpoints in the study. These covariates will be used in efficacy analyses to increase the precision of treatment effect estimates. The sample size calculation for the study will account for the expected increase in precision from the covariate to lower the required sample size, and will be informed by previous clinical trials, observational studies or registries, or real-world data. Robust estimation of other factors influencing the sample size calculation will be used to ensure adequate powering of the study, independent of the prognostic covariates. |

| **Step 3: The Risk of the AI Model** |
| --- |
| The decision consequence of the question of interest is low, as a study with lower power does not affect patient safety, type I error rate, or the validity of evidence from the trial. Adverse outcomes for the question of interest – a lower-powered study – is mainly a sponsor risk as it leads to a higher risk for an effective treatment not showing a statistically significant treatment effect.<br><br>The model influence is complex to characterize. The risk for reducing power, and hence the overall level of model influence, is heavily dependent on the amount of sample size reduction planned.<br><br>A small amount of sample size reduction with PROCOVA can minimize the risk of losing power and lead to a low model influence. A large amount of sample size reduction can lead to a large model influence. There are other factors, such as the expected endpoint variance and the dropout rate, that if mis-estimated can lead to a lower-powered study. If these factors are uncertain, it can lower the model influence. |

The following recommendations for risk mitigation should be applied with consideration for a specific use case:

1. **Model Evaluation**
    a. Conduct model evaluation as thoroughly as possible, prioritizing relevant data for testing as well as all endpoints relevant for powering. Use well-defined metrics, such as the correlation coefficient or the variance reduction achieved in the analysis of a study or through a bootstrapping procedure to mimic a study such as randomization-based inference. If planning a phase 3 trial, evaluation of performance of the model-derived covariate on the preceding study (e.g., a phase 2) is ideal, especially if trials are very similar in population (inclusion/exclusion criteria, region, and recency). Evaluation for the complete set of relevant endpoints is important, including key secondaries that influence sample size and power considerations.
    b. Multiple test cases are ideal. It is recommended to hold out any test data from the training procedure so that the model is not optimized on data being used for testing. Nested cross validation performance is equivalently valid and can be divided by study or other means to form multiple test sets. In cases of limited relevant data for evaluation, or as supplementary evidence, cross validation performance is informative but should be discounted.
    c. Performance testing is ideally conducted with the same baseline variables and endpoints that are expected to be measured in the planned study. If a baseline variable is present in the model but will not be measured in the planned study, it should be ablated from test data (marked missing), or the model should be trained to not include it (see Step 4 on model development).
    d. Use a conservative estimate of the expected variance reduction to limit the risk of losing power. For example, apply a fixed factor when computing sample size, such as 90% of the expected variance reduction. Or, given the target sample size of the study, bootstrap the expected variance reduction to estimate the finite sample size variability and take a fixed percentile such as 20% when computing the sample size. These conservative factors are useful when the sample size reduction is large to limit risks from losing power if other strategies do not provide sufficient risk mitigation.
2. **Blinded interim sample size re-estimation**
    a. When an interim analysis will be conducted, a blinded sample size re-estimation can significantly mitigate the risk of reducing power. The reduction of risk depends on the size of the interim analysis and timepoints evaluated. Simulation studies should be conducted to support conclusions on the risk of reducing power and sample size re-estimation.
3. **Consideration for prognostic information in standard covariates and stratification factors**
    a. When standard covariates or stratification factors are also expected to be meaningfully prognostic, they can provide a protection against reducing power. Sponsors should define whether the target power is for an analysis with no covariates or with adjustment for standard covariates or stratification factors. If in the planned study the model-derived covariate is less prognostic than expected, the added prognostic value of the standard covariates and stratification factors can add precision and limit power reduction.

The totality of testing results should be assessed when selecting the magnitude of sample size reduction, in concert with the other risk mitigation strategies below. No specific algorithm is recommended here to define the sample size reduction, but the following approaches are recommended:

- List all testing results across studies/cohorts and endpoints, with a simple grading of relevance that considers recency and population similarity and any important considerations.
- List any other risk mitigation strategies employed and the expected benefit of them to limit reducing power. The impact of both a blinded sample size re-estimation and the prognostic information contained in standard covariates and stratification factors should be described quantitatively.
- List any other factors that may influence the level of risk tolerance that the sponsor has for reducing power. This may include enrollment challenges or ethical motivations to expose fewer participants to placebo.
- List other factors that will govern the sample size determination, such as regulatory requirements or sponsor goals for safety data collection, or the need to maintain a randomization ratio for study design considerations.
- Select a sample size reduction based on the above data, and how the sample size will be reduced (e.g., maintaining randomization ratio or reducing the control arm). Provide a written justification of the sample size reduction. Provide this information as part of Steps 6 and 7 of the guidance framework.

**Step 4: Credibility Assessment Plan**

There are several recommendations for data, modeling, and evaluation specific to this context of use:

- **Data**
    - Broad, relevant datasets are generally recommended for PROCOVA applications. As discussed below, generalization performance is paramount, and for most machine learning architectures the exposure to more, relevant data increases generalization performance. Providing the model the opportunity to learn from patterns in variables on data is valuable for generalization, even if the data has a different severity, slightly different feature set available, is older, or has reasonable covariate shifts. The degree of relevance is subjective, and experimentation is usually warranted to evaluate impacts on generalization. In general, evaluation of data for population drift is important to gauge relevance for future trials.
    - Only features planned to be used during the study should be implemented in the model. Other features may be relevant to understanding the disease course or even inform which baseline data is taken in the study, and hence may be of interest for data collection. For the model itself, the input features should be a subset of the intended data collection at baseline of the study.
        - In some cases, such as if the model has been already trained and validated, features may exist in the model that are not collected in the study. Previous testing may motivate not updating the model. In these cases, it is recommended to evaluate the sensitivity of the model to the features that will be missing in the study to understand impacts on performance.
    - Data harmonization across data sources is important to establish a consistent data input

into the model. Care should be taken to ensure data sources used for training have a consistent representation (e.g., distributions are similar and differences understood), and that a defined data specification is used for prediction to ensure that the inputs to the model are consistent. Data quality is important in building a training dataset but vital in model evaluation and use, as unidentified data quality issues can skew model performance.

- **Model development**
  - There are many ways to build a model to predict standard of care outcomes, and no single architecture or approach is recommended as the needs and constraints may vary by indication and endpoint. For clinical trials with multiple endpoints of interest, multi-outcome models such as digital twin generators are useful to capture relationships between endpoints and make modeling efficient and manageable, as well as to enable related applications.
  - Generalization of prognostic performance to the planned trial is the most important property of model performance. The prognostic performance of the model-derived covariate must remain high for the intended study. Therefore, it is recommended to consider architectures and training methods that support good generalization performance. Regularization may be needed, and monitoring of training and tuning losses during training is recommended.
  - Overfitting is a concern if the model complexity or expressiveness is high. Overfitting can directly act against generalization, but expressive models may be required to capture subtle conditional relationships in the data. Therefore, training procedures to mitigate overfitting, such as using cross validation or nested cross validation, are recommended. Nested cross validation has the added benefit of generating held out model performance metrics for all data in the outer folds, increasing the breadth of model evaluation metrics and allowing for a clearer evaluation of model credibility.
  - Missing data may be a challenge if training data is aggregated across sources and some sources do not collect all features used in the model. Pre-imputation or co-training of imputation models may be used to handle missing data in the training dataset and at inference time (when missing baseline data must be imputed).
- **Evaluation**
  - The recommended evaluation metric is the variance of the treatment effect estimate for different covariate sets (none, standard covariates, model-derived covariate, standard covariates and the model-derived covariate), and the variance reduction compared to a fixed reference such as no adjustment. These can be quantified through the Pearson correlation coefficient between the model-derived covariate and the observed endpoint for a given cohort, which is asymptotically related to variance reduction (variance reduction is the square of the correlation), or it can be evaluated through statistical analysis models either using known treatment assignments or randomization-based inference.
    - While accuracy of the model-derived covariate for prediction of standard of care outcomes is of interest, it is irrelevant for the actual context of use. Because model predictions are being used as a covariate, the accuracy of the covariate is not relevant to the impact on the treatment effect estimate or precision. For example, we could scale any covariate by a constant factor, and it would not

change the treatment effect estimate.
  - Evaluation should be performed in cohorts similar to the intended study, or cohorts similar to related study designs. Cohorts that are too broad risk over-inflating variance reduction metrics due to increased heterogeneity that is more easily captured by the model.
    - One risk mitigation strategy to consider as part of evaluation is to compute conservative forms of the evaluation metrics for the planned trial. This is described in the box in Section 3.
  - A larger evaluation dataset (ideally across multiple studies or cohorts) provides a more robust evaluation of performance of the model-derived covariate. Sample sizes should be considered, as small cohorts have large statistical uncertainty and may be uninformative.

**Step 5: Executing the Credibility Plan**

There are several best practice recommendations for *carrying out* the steps in the credibility plan.

- **Data integrity**
  - Datasets should be reproducible and traceable to the original source data.
  - Code that produces datasets should be versioned, and data specifications recorded that specify important data properties (e.g., type, allowed range or values, definition).
  - Data used for training and testing should be separated in software to ensure data leakage cannot occur (e.g. training on test data), and software testing should be used to check for data leakage.
  - Any data transformations used in model training should only be determined from training data.
  - Pre-randomization data used as input to the model for testing should not contain any forward-looking (post-randomization) information.
- **Model training systems**
  - Model training should be reproducible, e.g. with random seeds fixed prospectively.
  - Trained models should be serialized (saved) at the end of training and should be able to be deserialized (loaded) for prediction.
  - Training logs should be stored as documentation of model training. Loss metrics, if used for model selection or key diagnostics, should be recorded.
  - Model code should be versioned and tested to ensure that data leakage constraints are respected (no leakage, no forward looking information in prediction).
- **Evaluation analyses**
  - Test data should be distinct from training data, with data pipelines ensuring there is no leakage (training data being part of test, or test data being a part of training).
  - Testing comprises two key steps: prediction and evaluation. The prediction step should not use forward-looking information beyond baseline, and more generally the test data should look as similar to planned use cases as possible. This means, for instance, removing features that are known to be missing in the future application if they are present in the model (ideally, they are not input features to the model). It also means that data transformations employed for training are appropriately applied to input data for testing so that predictions are consistent.

- Predictions should be saved before evaluation so that evaluation is reproducible and can be performed later.
- Evaluation should use data and transformation rules that are as identical as possible to the planned use case. For example, imputation rules applied to outcomes data in the planned study should also be applied to the outcomes data in the test data.
- If evaluating model-derived covariate performance using one or more complete studies, the testing should be performed in two ways: (1) each arm should be evaluated independently using asymptotic performance metrics (e.g., the correlation between the model-derived covariate and the endpoint) and (2) with the planned statistical analysis for the study. The first approach provides a consistently interpretable performance metric, while the second provides the comparison that would be expected in a study.

**Step 6: Documenting the Results**

The following tables should be included in the documented results of the credibility assessment plan.

- **Evaluation results**
  - A table of variance reduction metrics. Per evaluation cohort and endpoint, the following quantities should be computed and reported:
    - The sample size
    - The variance reduction of adjustment with standard covariates over no adjustment, as a percentage
    - The variance reduction of adjustment with the model-derived covariate plus standard covariates over no adjustment, as a percentage
    - The variance reduction of adjustment with the model-derived covariate plus standard covariates over standard covariates adjustment, as a percentage (note this quantity is not a simple difference of the previous two)
  - Metrics should be computed using a consistent method across cohorts, e.g. using the correlation or partial correlation and its relation to the asymptotic variance reduction, $R^2$. If cohorts comprise complete studies then statistical analyses may be used, or randomization-based inference with statistical analyses if non-study cohorts are used. One should consider, however, whether finite sample effects or other data features that influence variance are interpretable when a more complex statistical analysis is used, such as MMRM.
- **Relevant design features**
  - Key trial design features and their influence on power, including:
    - Sample size and power of the design without the impact of the model-derived covariates in the power calculation. Power should be reported for each endpoint of interest considered in the sample size calculation (e.g., the primary and key secondaries).
    - Goals for power when using sample size reduction, e.g. to maintain the design power or the power achieved with the inclusion of standard covariates in the analysis. This defines the reference point for assessing benefits of the model-derived covariate.
    - Statistical analysis methods planned for each endpoint.
    - Parameters of an interim where sample size re-estimation is planned, if using.

| | This includes the conditions for the interim (e.g., fraction enrolled or completed), as well as quantitative measures to mitigate the risk of losing power. |
|---|---|

| **Step 7: Evaluating Adequacy for the Context of Use** |
|---|
| For a given amount of sample size reduction, two considerations should be weighed:<br>1. The degree of risk for reducing power<br>    a. This risk can be largely quantified, for example the minimum power expected if the model-derived covariate provides no prognostic value or if it provides a minimum level of prognostic value motivated by evaluation performance and design features of the study.<br>2. The meaningfulness of the amount of sample size reduction<br>    a. There may be a minimum amount of sample size reduction above which it is worth the added logistical complexity or cost.<br><br>The sample size reduction chosen should balance these considerations. First, it should be determined if meaningful sample size reduction can be supported within tolerable risk levels for reducing power. There may be circumstances where the sample size reductions that can safely be realized are insufficient for the added complexity. Second, the amount of sample size reduction should be chosen, and a written evaluation should be made of the risk of reducing power, both in terms of likelihood and degree. This risk presentation should be expected to be a central point of discussion with regulators. |